\documentclass[sn-chicago]{sn-jnl}

\usepackage{graphicx}%
\usepackage{multirow}%
\usepackage{amsmath,amssymb,amsfonts}%
\usepackage{amsthm}%
\usepackage{mathrsfs}%
\usepackage[title]{appendix}%
\usepackage{textcomp}%
\usepackage{manyfoot}%
\usepackage{booktabs}%
\usepackage{algorithm}%
\usepackage{algorithmicx}%
\usepackage{algpseudocode}%
\usepackage{listings}%

\usepackage[utf8]{inputenc}
\usepackage{color,soul}

\usepackage[table]{xcolor}
\definecolor{red}{HTML}{9fc5e8}
\definecolor{Gray}{gray}{0.9}

\begin{document}

\title[The Best Ends by the Best Means: Ethical Concerns in App Reviews]{The Best Ends by the Best Means:\\Ethical Concerns in App Reviews}

\author{\fnm{Neelam} \sur{Tjikhoeri}}\email{n.tjikhoeri@student.vu.nl}
 \equalcont{These authors contributed equally to this work.}

\author{\fnm{Lauren} \sur{Olson}}\email{l.a.olson@vu.nl}
 \equalcont{Authors contributed equally to this work.}

\author*{\fnm{Emitzá} \sur{Guzmán}}\email{e.guzmanortega@vu.nl}

\affil*{\orgdiv{Faculty of Science}, \orgname{Vrije Universiteit Amsterdam}, \orgaddress{\street{De Boelelaan 1111}, 
\city{Amsterdam}, \postcode{1081 HV Amsterdam}, \country{The Netherlands}}}


\abstract{
This work analyzes ethical concerns found in users' app store reviews. We performed this study because ethical concerns in mobile applications (apps) are widespread, pose severe threats to end users and society, and lack systematic analysis and methods for detection and classification. In addition, app store reviews allow practitioners to collect users' perspectives, crucial for identifying software flaws, from a geographically distributed and large-scale audience. For our analysis, we collected five million user reviews, developed a set of ethical concerns representative of user preferences, and manually labeled a sample of these reviews. We found that (1) users highly report ethical concerns about censorship, identity theft, and safety (2) user reviews with ethical concerns are longer, more popular, and lowly rated, and (3) there is high automation potential for the classification and filtering of these reviews. Our results highlight the relevance of using app store reviews for the systematic consideration of ethical concerns during software evolution.
} 



\maketitle
\section{Introduction}
Ethical concerns like deteriorating mental health, discrimination, privacy, and manipulation are standard in today's software products (\cite{2nytimeseatingdisorder,1facebookfiles2021,3zuboff,5reallifemag,4gebru2019oxford,7nytimeswrongaccused,6noble2018algorithms,10guardianscanned,8kindletrack,9whatsapp,11vicedata,12guardianbrexit,13roy2017cathy}). These examples are just a subset of users' ethical concerns regarding modern software. In 2023, over 85\% of the human population owns a smartphone, full of software applications crafted by an exploitative monopoly of mostly US-based corporations (\cite{abuseproject2022, mergeproject2022,statista}). Users' ethical concerns about software have increased as the ability to regulate software technologies decreases. Regulation is difficult because of software platforms' centralized control over technology and the inextricable embedding of modern technologies in daily human life. 

A few studies have examined singular ethical concerns expressed in user feedback from social media and app stores, as these sources are effective for identifying software deficiencies from a massive, dispersed, and highly diverse user base. 
This recent research has investigated singular ethical concerns like privacy (\cite{besmer2020investigating, li2022narratives}) and discrimination (\cite{tushev2020digital}), and demonstrates the potential in revealing users' perspectives on ethical concerns (\cite{besmer2020investigating, tushev2020digital}) and connecting users' ethical concerns to real-world events (\cite{li2022narratives}). 

Our work goes a step beyond these existing studies by putting together a broader \textit{set} of ethical concerns about software mentioned in literature and user feedback, and also examining them through manual and automatic analysis. Developers can use this developed set of users' ethical requirements during software design and evolution, and identify new ethical concerns within their current software. These concerns can serve as a checklist for practitioners before software release. Similarly, the manual and automatic analysis can serve as a reference for practitioners seeking to incorporate the consideration of users' ethical concerns into their software processes. 

In our work, we first created a set of ethical concerns based on Wright's ethical values framework (\cite{wright2011framework}). Next, we collected 5,864,188 user reviews and manually labeled 3,101 using our set of ethical concerns. During the annotation, we supplemented the set of ethical concerns to reflect users' current concerns outside the existing literature. Then, we considered the frequency of ethical concerns and their co-occurrence with different applications. Finally, we used the annotated data for training and testing predictive models to automatically detect ethical concerns and classify them into the different categories of our set of ethical concerns. 

 Currently, there is no systematic method to identify unaddressed ethical concerns during software development (\cite{grundy2021impact}), and research has not yet analyzed the extent of ethical concerns in existing software applications. This work is a critical first step towards developing a \textbf{systematic approach for considering users' ethical concerns} during software development. Our accessible, widely applicable methodology will allow software companies and development teams to systematically and proactively consider ethical concerns, manage risk, and comply with legal requirements.

This research's main contributions to the field of Empirical Software Engineering (ESE) are:

(1) A qualitative and quantitative analysis of 16 ethical concerns encompassing users' perspectives on software. This analysis provides insight into current unaddressed ethical requirements about software held by users across multiple domains. Within this analysis, we establish patterns within our found ethical concerns which can apply to all software products. 

(2) A manually annotated dataset of users' ethical concerns, and machine learning models effective at identifying ethical concerns within filtered data. Practitioners can use this data and models to incorporate users' ethical concerns into software evolution.

\begin{figure*}[h!]
    \caption{Set of Ethical Concerns and Modification of Wright's Framework}

    \centerline{{\includegraphics[width=1\textwidth]{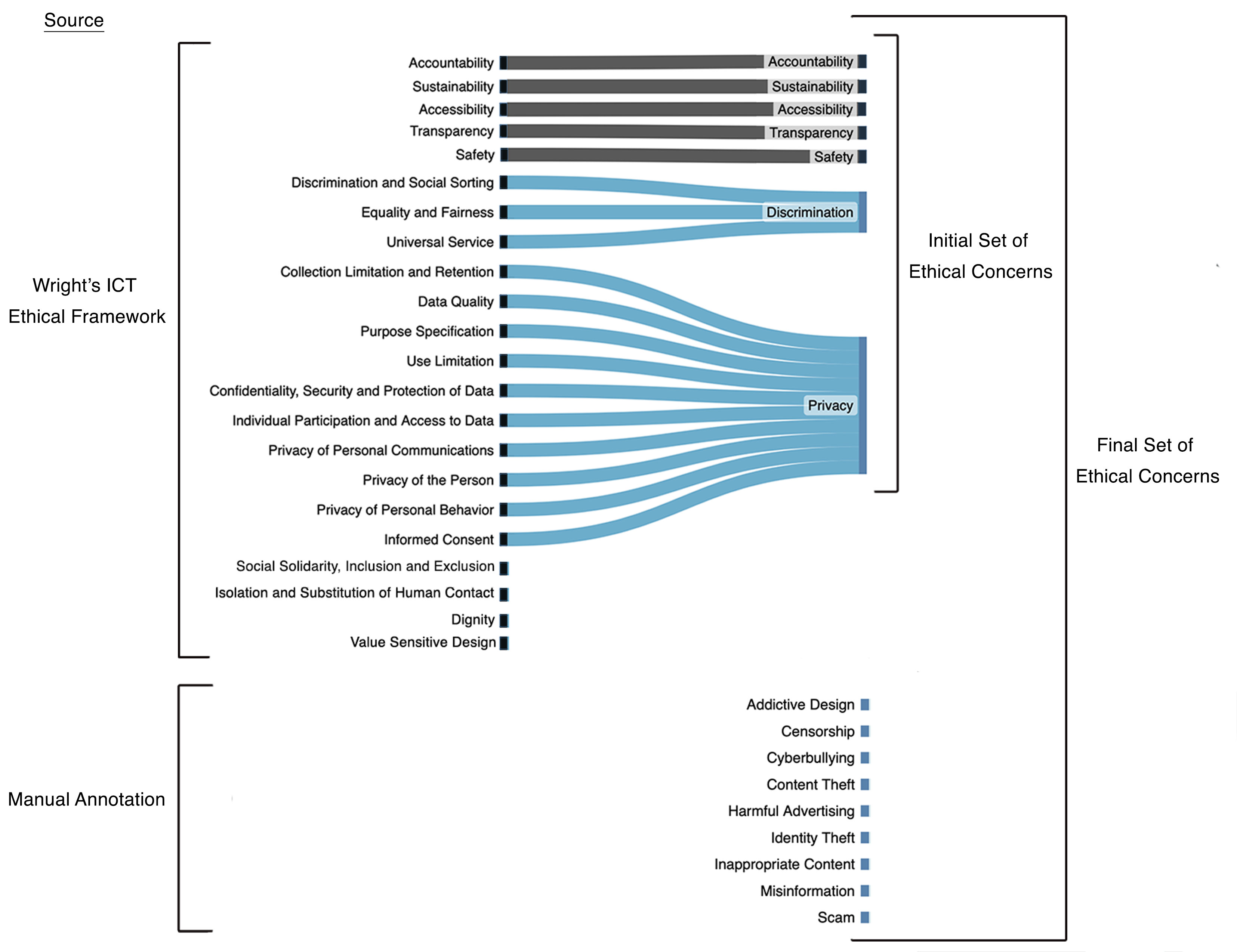}}}%
    
    \label{fig:taxonomy_mods}
\end{figure*}
\section{Research Method}
We focus our study on three research questions:

\textbf{(RQ1)} Which \textbf{types of ethical concerns} are present in user reviews and how do they change over time?

\textbf{(RQ2)} What are the main \textbf{characteristics} of user reviews that express ethical concerns?

\textbf{(RQ3)} What is the \textbf{automation potential} of identifying and classifying ethical concerns from a large set of user reviews?

While there are several channels in which users write reviews about the software they use, we chose to focus on app stores because they are the most studied channel to date. We specifically selected the Google Play Store because it is the largest app store on the market (\cite{statista}). 

To answer our research questions, we first establish an \textbf{initial set of ethical concerns} based on the literature. Then, we \textbf{collect data} from the Google Play Store, from various applications across domains. After, we \textbf{sampled these data} and through trial rounds of manual annotation, we collected both (1) a \textit{set of keywords} to filter for ethical concerns and (2) a \textit{set of additional ethical concerns}. Then, we \textbf{manually annotate} a sample of 3,101 posts to identify and classify ethical concerns within these user reviews. Finally, to prepare for the process of \textbf{automatically analyzing} ethical concerns, we trained a binary and multiclass classifier with our previously annotated data. In the following, we describe each of these steps in more detail. 

\subsection{Initial Set of Ethical Concerns}\label{sec:taxonomy}
To categorize ethical concerns found in user reviews, we use Wright's Information and Computer Technology (ICT) Ethical Impact Assessment Framework as a starting point (\cite{wright2011framework}). We chose this framework in contrast to other existing sets of ethical concerns as it focuses on ICT as a whole rather than too restricted of a domain like just AI or too broad of a domain like all possible ethical concerns. However, we found that some values from Wright's framework would be too academic to appear in user reviews. Because of this, we simplified some of Wright's values to reflect more vernacular ideas and language.

The values that remain unchanged from Wright's framework to our set are sustainability, safety, accessibility, accountability, and transparency. These values were at the right level of complexity to align well with user perceptions of ethical concerns of mobile applications. For ethical concerns relating to the proper maintenance of user data, Wright had multiple more fine-grained categories. We merged these into one larger privacy category. Wright also had multiple ethical values related to discrimination, which we combined into one category. We excluded four concerns mentioned in our reference literature (\cite{wright2011framework})---value sensitive design, social solidarity, inclusion and exclusion, dignity and isolation, and substitution of human contact---from our ethical concerns set. Although these critical values warrant focus, we posit that the concepts behind them are too abstract and academic to be found in user reviews.  Although Wright's framework covers the entire ICT domain, Wright created this framework in 2011, before the rise of the current technological landscape, which creates a gap in our initial set of concerns. To address this gap, we add categories to Wright's framework from ethical concerns expressed by current users, as detailed in Section \ref{sec:annotation}. Figure~\ref{fig:taxonomy_mods} shows Wright's set of ethical concerns, our modifications to his set, and our set of ethical concerns before and after adding new categories from manual annotation.

\subsubsection{Internal and External Concerns}
\label{sec:inex}
Not only has mobile application software created ethical concerns for users in virtual spaces, but these concerns have also bled into the physical realm. Within the past few years, software companies have received criticism for influencing elections, inciting violence in the real world, and facilitating sexual assault (\cite{brattberg_maurer_2018, guardian_2018, raitanen2019deep,lee2022storm,o2018cnn}).
To identify and analyze these material matters, we consider any concerns that involve the physical world as \textit{external} ethical concerns. We identified six ethical concerns as potentially external: accountability, discrimination, safety, scam, sustainability, and accessibility\footnote{The scam ethical concern was added after the analysis in Section~\ref{sec:annotation}}. This consideration can help developers understand the scope of the ethical concern and help them identify and mitigate it. Similarly, we consider \textit{internal} ethical concerns as those that occur within the realm of the software application. For example, end users discriminate against end users within the software platform.  

\subsection{Data Collection}
We collected data from Alexa, Facebook, Google Home, Instagram, LinkedIn, TikTok, Vinted, Uber, YouTube, and Zoom. We chose these applications due to their popularity, differing functionalities, and large and diverse target audience. We consulted App Annie (now data.ai\footnote{https://www.data.ai/en/}), a leading platform for app analytics data, for the top apps within our country of residence. All the apps included in our analysis come from this list. We chose apps from different domains, assuming that different target audiences would use different applications, and that ethical concerns might highly depend on each domain. We removed games from consideration because previous research shows that the nature of reviews within this domain differs significantly~\citep{Guzman2014}. 

We next gathered data using an existing scraper (\cite{jomingyu}). We collected a maximum of one million reviews for each application. However, some applications did not have one million reviews (Zoom, LinkedIn, Google Home, Alexa, Vinted); we scraped all the available reviews for those applications.

\subsection{Data Sampling}
For the nine trial rounds of annotation, each trial consisted of five randomly sampled reviews for each of the ten applications, totaling 50 reviews. For our first sample of 50 reviews, we found that only 2\% contained any ethical concerns. Because of this extreme scarcity, we decided to do an additional keyword-based filtering process to ensure that we could generate a large sample of users' ethical concerns. As we sampled reviews, we iteratively gathered keywords relevant to ethical concerns. Through this filtering, we increased the prevalence of ethical concerns to 30\% within our last trial sample. Section~\ref{sec:keywords} gives more information on the used keywords and their selection.  For our final sample, we randomly selected 200 reviews for each application using the keyword list. This selection resulted in a sample of 2000 reviews for the ten applications. 

\subsection{Manual Annotation Setup (RQ1)}
\label{sec:annotation}
\begin{figure*}[h!]
    \caption{Manual Annotation Process}

    \centerline{{\includegraphics[width=1\textwidth]{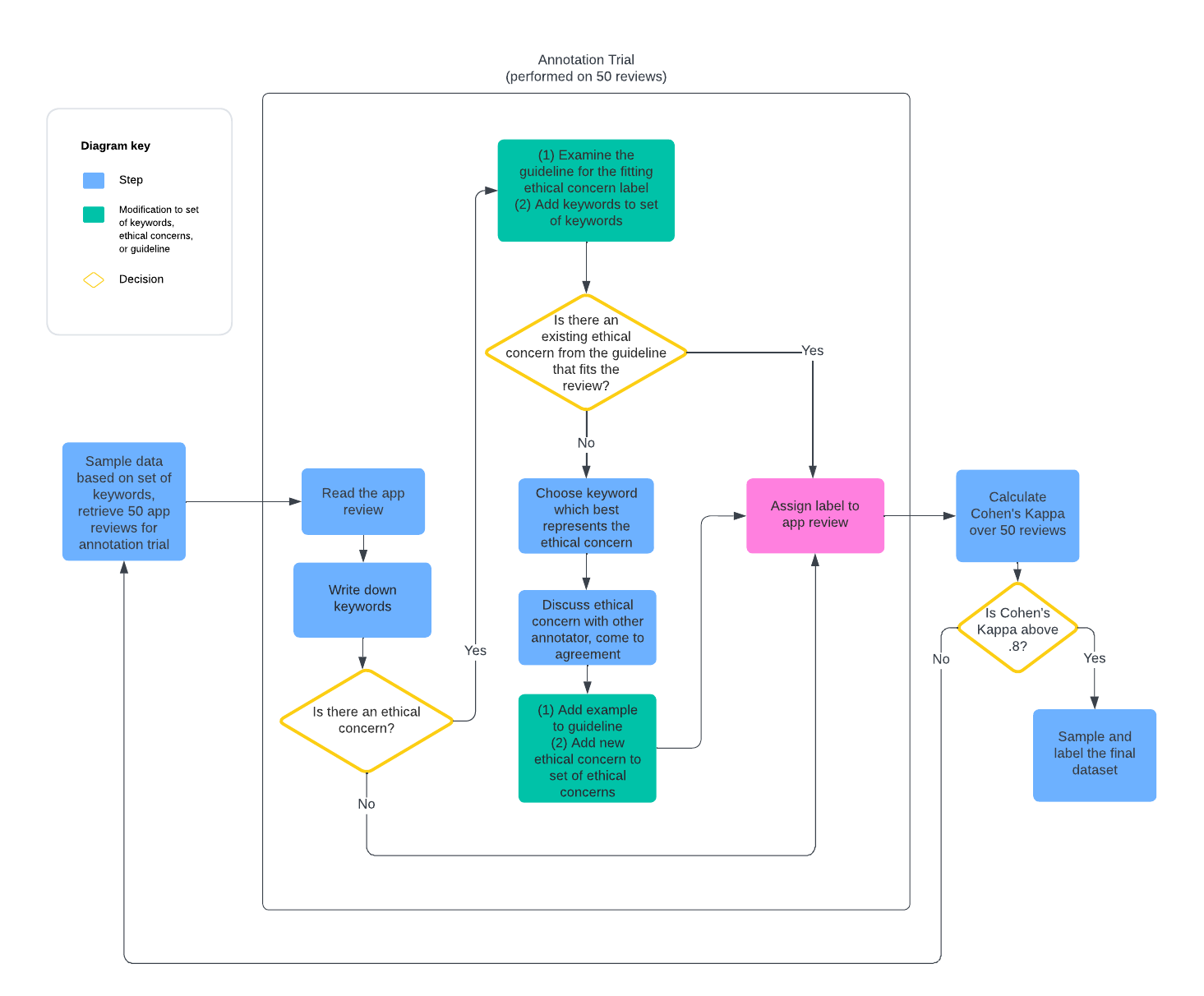}}}%
    
    \label{fig:annotation_process}
\end{figure*}

Figure~\ref{fig:annotation_process} shows an overview of the manual annotation process. Annotators labeled the reviews with the assistance of a guideline, included in our replication package\footnotemark[3], which provided a list of initial ethical concerns, its definitions and examples. This initial list was extended throughout the trials. The \textit{initial set of ethical concerns} are shown in Figure~\ref{fig:taxonomy_mods}. Annotators only labeled the \textit{negative mentions} of an ethical concern in reviews (e.g., ``The privacy settings in this app are insufficient") or when there is \textit{clearly something harmful} described in the review even though this can be presented in a positive way by the user, like references to app addiction (e.g.,``Love this app so much that I'm using it 24/7"). If there was more than one ethical concern present in a user review, they were ordered by importance\footnote{We only considered the first ethical concern in subsequent analysis}. Internal and external labels were also given to each ethical concern (see Section~\ref{sec:taxonomy}). 

 To ensure the quality of the annotations, both annotators conducted nine annotation trials before annotating the final dataset. With each trial, \textit{new filtering keywords} were added to obtain a sample with a higher concentration of ethical concerns as well as a broad range of ethical concerns accurately encapsulating phenomena described by users. There was also notice taken to potential \textit{new ethical concerns} that were not yet in our initial set of ethical concerns. Whenever we found such a concern, both annotators discussed the new concern. If both annotators agreed, we added it to our set of ethical concerns. Snowballing for keywords has been used in  other research analyzing ethical concerns in user feedback to overcome sparse data~\citep{tushev2020digital, obie2021first}.  

\subsubsection{Final Set of Keywords}\label{sec:keywords}
We applied keyword filtering to ensure that some of the analyzed reviews contained information relevant to ethical concerns in software applications. In our first sample, we used a stemmed version of the initial set of ethical concerns as well as their synonyms, found through WordNet, giving us 20 keywords as a starting point. In each trial round of annotation, we added relevant keywords through a snowballing approach (\cite{tushev2020digital}). The total number of keywords in the final list is 109, found through nine trial annotations of 50 reviews each. The keywords used for filtering the dataset that was later manually annotated are available in our replication package\footnotemark[1].


\subsubsection{Final Set of Ethical Concerns}
\label{sec:mods}
Whenever annotators found an ethical concern from users that did not fit into the existing set of concerns, annotators discussed adding an additional concern. If both annotators agreed that this concern was a new phenomenon uncovered by current definitions, we added it to our set of concerns. We found nine additional concerns: addictive design, cyberbullying, censorship, content theft, harmful advertising, identity theft, inappropriate content, misinformation, and scam. The bracket on the bottom left of Figure~\ref{fig:taxonomy_mods} indicates all the ethical concerns added during the manual annotation. Table~\ref{table:taxonomy} shows our final set of ethical concerns, and their definitions, while Table~\ref{table:some_reviews} shows samples of user reviews which reference these ethical concerns. 


\begin{table*}[h!]
\centering
\caption{Ethical Concerns' Definitions}
\begin{tabular}{p{0.2\textwidth}|p{0.8\textwidth}}
\hline
\textbf{Ethical Concern}  & \textbf{Definition}  \\\hline

\rowcolor{Gray}Accessibility & The application does not include people with special needs or disabilities. This concern can be about the usage of the application itself or about a service the application is offering.\\

Accountability & The user experienced an issue when using the application or its service. The user could not find the software company responsible for solving the issue.\\

\rowcolor{Gray}Addiction & The user mentions how they are addicted to the application or describes that they use it excessively.\\

Censorship & The application deliberately hides certain information, or certain users' content or profiles are deliberately removed or demoted.\\

\rowcolor{Gray}Content Theft & Content from a user is stolen or used without permission from the original creator.\\

Cyberbullying & The platform's community is being harmful, abusive, or unhealthy by practicing hateful communication via the application.\\

\rowcolor{Gray}Discrimination & The application user is being discriminated against by the application, its services, or its community. This concern also includes users who have an issue with having fewer functionalities available to them because they live in a different geographical area.\\

Sustainability & The user says something about the negative impact the application has on the environment.\\

\rowcolor{Gray}Harmful Advertising & The user notices the presence of deceiving, misleading, or harmful advertisements throughout the application.\\

Identity Theft & Someone is using the identity of someone else on this application. This concern also applies to catfishing, creating fake profiles to trick and deceive other users on the platform. \\

\rowcolor{Gray}Inappropriate Content & The application contains content other than advertisements that are disturbing to certain groups of people.\\

Privacy & The users' identity and data are not kept secure or used for purposes other than what the user gave consent to. This concern also includes when an account is hacked.\\

\rowcolor{Gray}Safety & The usage of this app has led to health issues or other safety risks. This concern can be about the usage of the application itself or its services.\\

Scam & The user has been scammed or came into contact with a scammer through the application. This concern can occur through the application only or its services. A scammer deceives another to gain something, usually money or goods.\\

\rowcolor{Gray}Misinformation & False information is spread through this application.\\

Transparency & The motives, risks, and implications are unclear to the user when using this application or a service the application provides.\\\hline

\end{tabular}
\label{table:taxonomy}
\end{table*}

\subsubsection{Additional Sample Creation} \label{sec:sampling2}
Our initial manually annotated data set resulted in a data distribution in which some ethical concerns were well represented, while others were present in minimal numbers. We noticed that popular ethical concerns such as privacy and censorship were found more frequently. To ensure that those with little representation were also included in our analysis, we conducted a second sampling round. 

In the second sampling, we focused on single ethical concerns rather than on all ethical concerns at once. Ethical concerns that were present with 5\% or less in the first sample were re-sampled to obtain more reviews reporting these concerns (all ethical concerns in the final set except for privacy and censorship). We sampled each ethical concern individually, with a limited keyword list specific to the particular ethical concern without considering differences among applications.

This strategy obtained more results for each ethical concern. 
Combining this second sample with our initial data set resulted in a final data set of 3101 reviews.

\subsubsection{Inter-Rater Agreement}
\label{sec:trial_runs}
The annotators continued to perform trials until establishing a high inter-rater agreement. The Cohen's kappa~(\cite{cohen1960coefficient}) between both annotators on the ninth (last) trial was 0.85, indicating a very good agreement~(\cite{coheninterpret}). The Cohen's kappa between both annotators for the final dataset is 0.91. The disagreements on the final data set were resolved through discussion on all points that presented dissent. 

\subsection{Analyzed Characteristics (RQ2)}
We consider three app review characteristics in our analysis: rating, word count and up-votes. These characteristics have been used in previous research to extract, classify and prioritize users' feedback~\citep{chen2014ar, panichella2016ardoc, villarroel2016release, guzman2015r, maalej2015r, guzman2017little}. We define them as follows.
\begin{itemize}
  \item \textbf{\textit{Rating:}} The score (from 1-5) the user gives to the app, represents the user's satisfaction with the app or ethical concern.  
  \item \textbf{\textit{Word Count:}} The number of words in the app review, represents the potential complexity and sentiment of a review or ethical concern.
  \item \textbf{\textit{Up-Vote:}} The number of `likes' an app review receives, represents the popularity of a review or ethical concern.
\end{itemize}

\subsection{Automated Analysis Setup (RQ3)}
We explored the use of traditional machine learning methods, as well as deep learning models, for the automatic extraction and classification of ethical concerns. We trained three types of predictive models: (1) binary classifiers for detecting the mention of ethical concerns, (2) binary classifiers for detecting whether an ethical concern is \textit{internal} or \textit{external}, (3) multi-class classifier categorizing the ethical concerns shown in Table~\ref{table:taxonomy}.  We use our data set of 3101 annotated reviews for training and evaluation of our models. We describe the methodology used below.

\textit{Preprocessing}
Since the data set consists of textual data, the data was first preprocessed using the NLTK~(\cite{documentation_2023}) library and then vectorized using TF-IDF as a weighting scheme. We also eliminated special characters and symbols, lowercased every character, tokenized, eliminated stopwords, and stemmed the resulting words.\\
\textit{Different Predictive Models}
 We tested five different models, namely Random Forest (RF), Multinomial Naive Bayes (MNB), Logistic Regression (LR), Multi-Layer Perceptron (MLP) and Support Vector Machine (SVM). We choose the Naive Bayes and SVM models, as they are recommended by Scikit-learn~(\cite{scikit}) for classification tasks with text data under 100k rows. We used LR, MLP, and RF as they are commonly used and perform well on a wide range of data. We trained and evaluated these models using Scikit-learn. 
 In addition, because of Bidirectional Encoder Representations from Transformers (BERT)'s high performance on a wide range of classification tasks, we implemented this deep learning model. We employed the BERT model through python's ktrain package~(\cite{Maiya_2023}) and utilized the "fit\_onecycle" policy with four epochs, a learning rate of 5e-05, and a batch size of 16 for training. The "fit\_onecycle" policy is an approach that adjusts the learning rate during training to optimize model performance.

\begin{figure}[h]
    \caption{Frequency of Ethical Concerns in the Final, Labeled Dataset}
    \centerline{{\includegraphics[width=.75\textwidth]{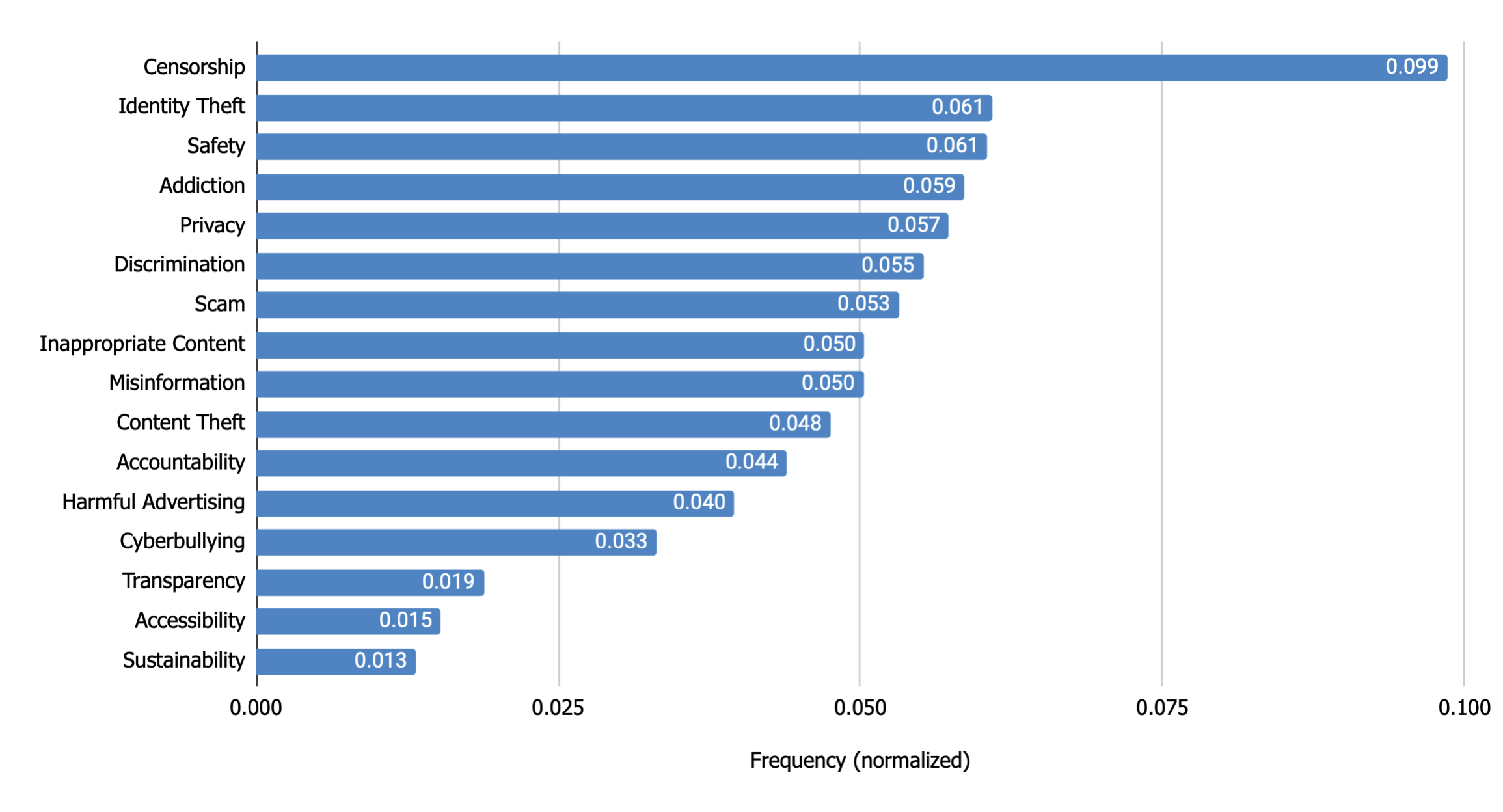}}}%

    \label{fig:distr}
\end{figure}

\textit{Handling Data Imbalance}
\label{sec:smote}
For binary classification tasks, there were more ethical concern reviews than nonethical concern reviews, signalling a slight data imbalance. Unfortunately, for multiclass tasks, there was a greater data imbalance, as shown in Figure~\ref{fig:distr}. We used SMOTE~(\cite{chawla2002smote}) to balance the training data and address this problem. Finally, just for the multiclass task, we removed ethical concerns with less than 50 reviews from the dataset used for prediction. The concerns whose predictions were excluded are harmful advertising, sustainability, accessibility, and content theft.

\textit{Evaluation Metrics}
We used a stratified 10-fold cross-validation to evaluate the traditional models and a split training-test data set of 80\% to 20\% for the BERT model.
 For all machine learning tasks within our study, we use the F1 score, precision, and recall as evaluation metrics. However, while we used the standard averaging method for reporting the results of the binary classifiers, we used \textit{macro} averaging, which treats all classes equally for the multiclass classifiers.

\section{Results}
\label{sec:res_manual}
\subsection{Types of Ethical Concerns (RQ1)}
\label{sec:qual}
In our manually annotated sample, 61.9\% (1919/3101) of the reviews expressed ethical concerns about the software. Figure~\ref{fig:distr} shows the frequency of ethical concerns by type within this data. To calculate frequency of ethical concern, each ethical concern's value per application was divided by the total reviews for that application, then averaged across all applications. This procedure scales the frequency so each application is balanced equally within the result, as there are an unequal amount of reviews per application in the final sample. Within our manually annotated sample, censorship, identity theft, and safety occur most frequently, while transparency, accessibility, and sustainability occur least frequently. Table~\ref{table:some_reviews} shows examples of labeled reviews for each ethical concern. 
 \begin{figure}[h!]
    \centering
    \caption{Frequency of Internal and External Labels Among Ethical Concerns}  
\includegraphics[width=.75\textwidth]{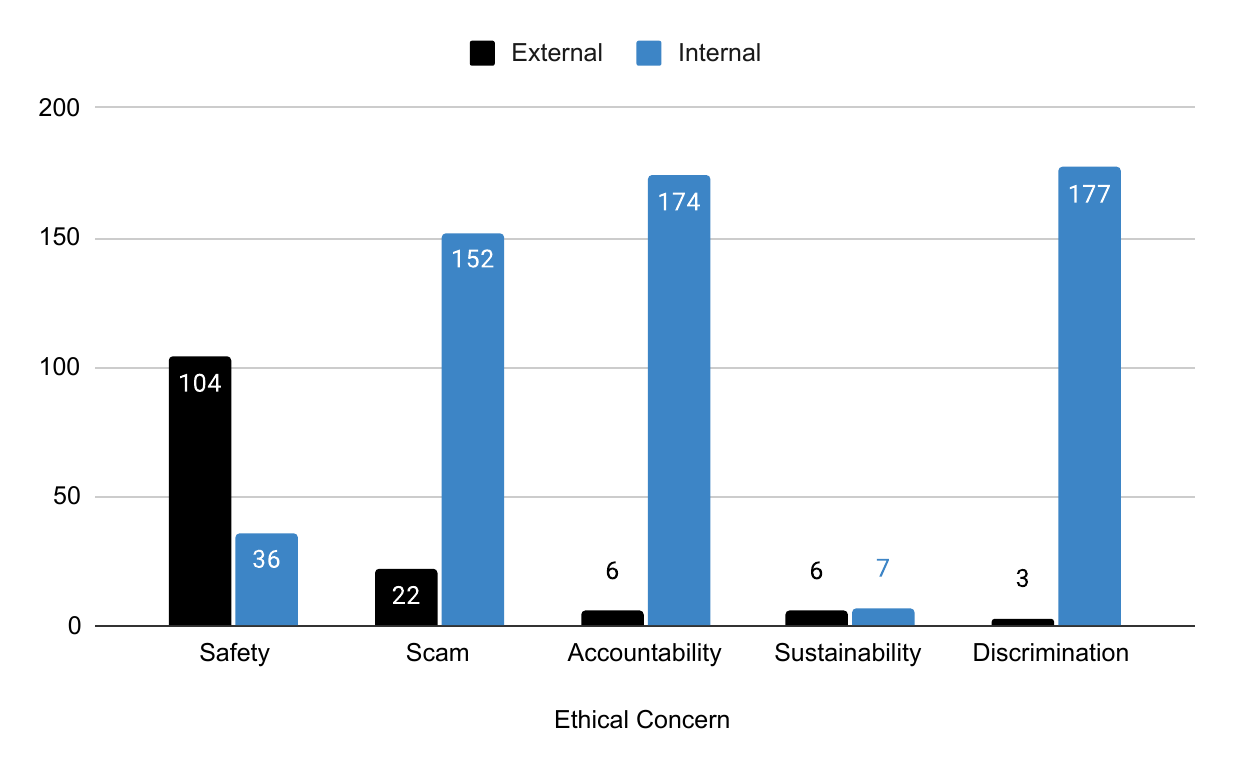}

    \label{fig:eth_inex}
\end{figure}

\textbf{Internal and External Ethical Concerns}
 Within the entire dataset, (1768/1919) 92\% of the ethical concerns were labeled \textit{internal}, meaning that there was an ethical issue within the application. Within the six categories we identified as potentially external (scam, safety, accountability, sustainability, discrimination), 80.3\% (583/726) of the ethical concerns were labeled \textit{internal}. Figure~\ref{fig:eth_inex} shows the breakdown by ethical concern. The two concerns with the most \textit{external} labels are scam and safety. These were mostly associated with the Uber application, with users reporting that drivers engage in sexual harassment and cause dangerous situations. See Table \ref{table:inex} for examples. 

\begin{table}[]
\caption{Examples of Internal and External User Reviews}
\begin{tabular}{|p{0.15\textwidth}|p{0.425\textwidth}|p{0.425\textwidth}|}
\hline
\small
\centering
\textbf{\textit{Ethical Concern}} & \textbf{External}  & \textbf{Internal} 
\\ \hline
\textit{Safety} & Awful customer service and they don't follow their own community guidelines. They allow dangerous medical misinformation and animal mistreatment to flourish. & I keep getting a message that the Alexa app has crashed without using the app. This is actually dangerous when I'm using a navigation app while driving and the Alexa message shows up and blocks my view of the screen so I can't see the map until I dismiss the Alexa message. \\\hline
\textit{Scam}            & After I started posting videos on TikTok, I began to receive scam phone calls.   &    It's difficult to use this app even though it should be easy. When I booked the ride today, the fare estimate was 39Rs but I was charged 100Rs.\\\hline

\textit{Accountability}  & Awful service. My Uber driver refused to come and did not answer my call. He didn't even cancel the ride. It was late at night and I was standing on road where there were no other options. Pathetic customer service...  &  I really like the app but unfortunately many creators aren't held accountable for violating community guidelines even after users report them.                        \\\hline
\textit{Sustainability}  & Uber should be made to pay an environmental emissions tax in every country it operates in without increasing prices or decreasing payouts to drivers.  & Since new update, the app is draining my phone battery where I have to charge the phone multiple times every day. I disabled the app and my battery is back to lasting all day. Please fix. Thank you.                                                                                                                                                                    \\\hline
\textit{Discrimination}  & The right wing has started using your partner drivers to spread communal poison (islamophobia, calls for violence) and verifiably fake news. Your app doesn't have a field where I can leave descriptive negative feedback, or upload conversation recordings, and that brought me here.  & I have never received a single job contact from this site, nor from this company that I have often applied too, I am sure like all the tech companies you do not hire African Americans, yet I still use the app because it's "professional".  \\        \hline                                                                                                                  
\end{tabular}
\label{table:inex}
\end{table}

\textbf{Extent of Ethical Concerns among Applications}
Amongst ethical concerns, discrimination, privacy, and transparency occur on \textit{all} platforms. 

\begin{figure}[h]
    \caption{Extent of Ethical Concerns among Applications}
    \centerline{{\includegraphics[width=.75\textwidth]{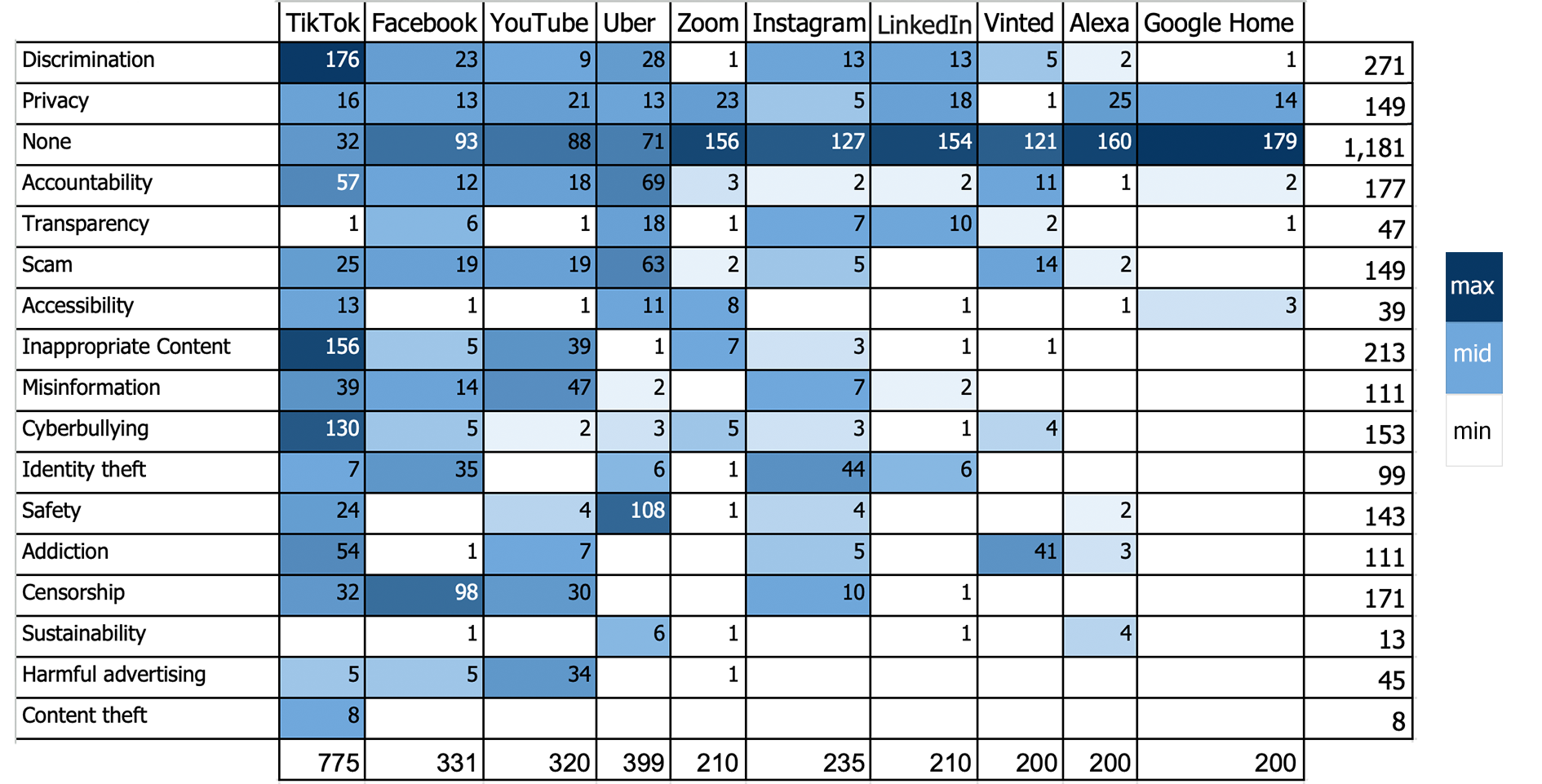}}}%

    \label{fig:ethicalconcerns-platforms}
\end{figure}


\begin{table*}[h!]
\caption{Examples of Ethical Concerns in User Reviews}
\centering
\begin{tabular}{p{0.17\textwidth} p{0.65\textwidth}  p{0.1\textwidth}}
\hline
\textbf{Ethical Concern}  & \textbf{User Review} & \textbf{Platform}  \\\hline

\rowcolor{Gray}Accessibility & It \textbf{doesn't let you put the subtitles (for Deaf viewers)} on TV when casting from the phone... & Google Home\\

Accountability & Calculate more fare than usual and \textbf{couldn't report for driver's bad behavior} & Uber\\
 
\rowcolor{Gray}Addictive Design & \textbf{
I really hate the YouTube rabbit hole and the design of YouTube as addictive. It is a very big problem.}  & YouTube\\

 Censorship & This application propagates radical ideology and \textbf{suppresses open-mindedness and conservative views}. I cannot express how much censorship is illegal and is what I observe on this platform.. &  Facebook\\

\rowcolor{Gray}Content Theft  & \textbf{Full of stolen content} that the creators worked for & TikTok\\

Cyberbullying & There's still alot of hate etc on there. \textbf{Its sad that you don't care about bullies} & Instagram\\

\rowcolor{Gray}Discrimination & Racist and discriminating. They have been \textbf{proven to suppress black lives matter, disabled, poor and anyone that's not conventionally attractive.} & TikTok\\

 Sustainability & \textbf{Imagine wasting power and contributing to carbon emission at such scale} where Facebook can avoid this by just introducing setting to invert colors.  &  Facebook\\

\rowcolor{Gray}Harmful Advertising & To many \textbf{advertising for gambling directed at kids} & Facebook\\

 Identity Theft & So far so good. \textbf{Lots of catfish. Men pretending to be woman and asking for money.}  & Instagram\\

\rowcolor{Gray}Inappropriate Content  & Theres \textbf{so many sexual videos} they might as well call this pornhub Junior & TikTok \\

Privacy  & The \textbf{amount of privacy you need to give up} to have this overhyped egg timer in your house is ridiculous. & Alexa\\

\rowcolor{Gray}Safety & Encourages young children do do \textbf{very dangerous tiktok `trends' like spraying your bathroom mirror with a flammable substance and setting it on fire.} & TikTok \\

 Scam &  Get \textbf{ scammers wanting money. And I have fell for a sob story.} Then the scammers delete their accounts & TikTok\\

\rowcolor{Gray}Misinformation & Amazing app it is my teacher but it just only one minor problem that \textbf{fake news is more than real news} & YouTube \\

Transparency & \textbf{Why does LinkedIn need to access calendar and confidential information from my phone?} New permissions are questionable & LinkedIn \\ \hline
\end{tabular}
\label{table:some_reviews}
\end{table*}


\label{sec:qa}
\begin{figure}[h]
    \caption{Frequency of Ethical Concerns per Application }
    \centerline{{\includegraphics[width=.75\textwidth]{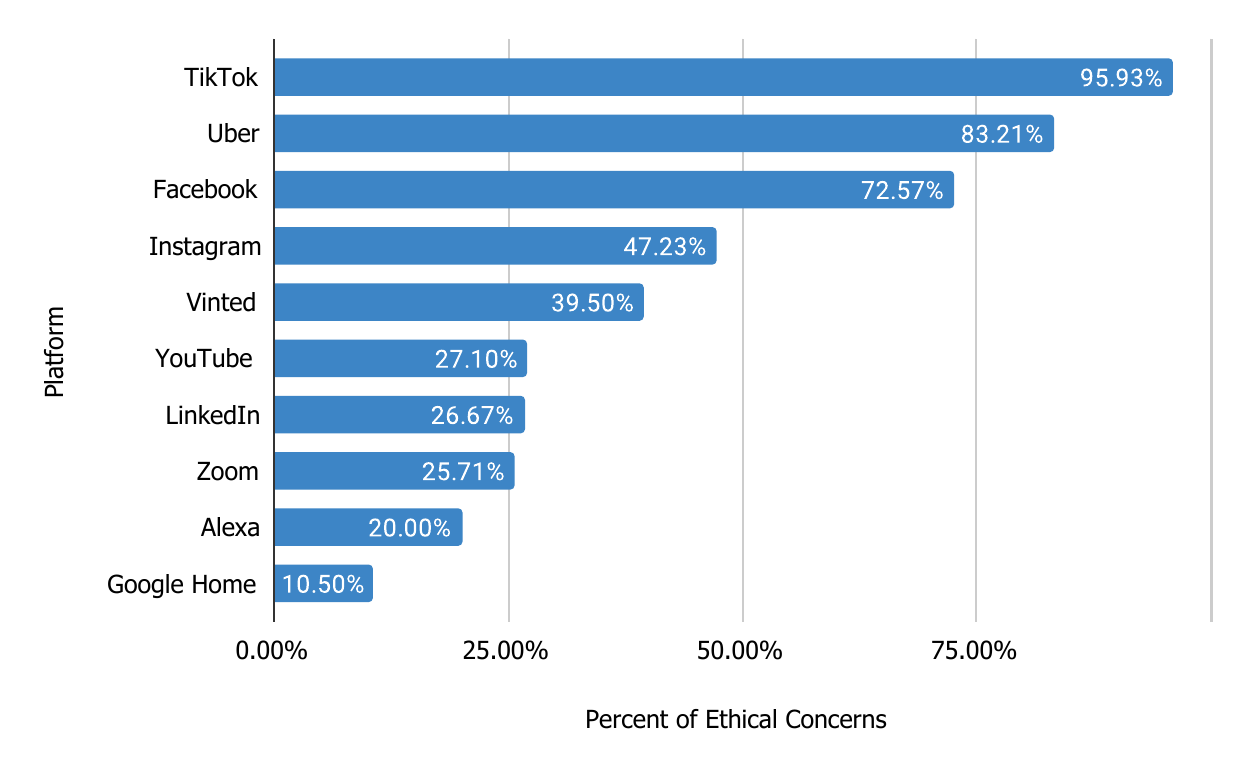}}}%

    \label{fig:app}
\end{figure}

\textbf{Ethical Concerns per Application}
As the design of applications heavily influences how ethical concerns manifest~(\cite{gray2018dark}), in this section, we inspect and detail the top three most frequently occurring ethical concerns per application. When considering this frequency, we take the number of ethical concerns reported on the app over the total amount of reviews analyzed for that app. Because we have different-sized samples per app, we do not compare based on the number of ethical concerns reported but on the percentage. Table~\ref{table:time} shows the time periods for each application. We posit that the reported ethical concerns per application regard these specific time periods.

\begin{table}[]
\caption{Temporal Data for Each Application}
\begin{tabular}{|l|r|r|r|r|}
\hline
\centering
\small
            & \textbf{Start Date} & \textbf{End Date} & \textbf{Time Period (years)} & \textbf{Mean Date} \\\hline
\textit{Zoom}        & 4/22/2019                      & 3/4/2021                     & 1.868                                   & 9/17/2020                               \\ \hline
\textit{TikTok}      & 7/15/2020                      & 3/20/2021                    & 0.680                                   & 11/23/2020                              \\\hline
\textit{Uber}        & 12/7/2013                      & 3/22/2021                    & 7.294                                   & 9/14/2018                               \\\hline
\textit{Facebook}    & 7/24/2020                      & 3/22/2021                    & 0.659                                   & 11/12/2020                              \\\hline
\textit{Instagram}   & 10/20/2020                     & 3/19/2021                    & 0.412                                   & 12/6/2020                               \\\hline
\textit{Vinted}      & 7/16/2013                      & 11/15/2020                   & 7.340                                   & 10/20/2015                              \\\hline
\textit{YouTube}     & 1/2/2021                       & 3/22/2021                    & 0.216                                   & 2/12/2021                               \\\hline
\textit{LinkedIn}    & 8/2/2011                       & 2/20/2021                    & 9.564                                   & 2/2/2017                                \\\hline
\textit{Alexa}       & 11/19/2015                     & 3/5/2021                     & 5.296                                   & 6/3/2019                                \\\hline
\textit{Google Home} & 12/3/2013                      & 2/12/2021                    & 7.200                                   & 7/6/2018      \\ \hline                         
  
\end{tabular}
\label{table:time}
\end{table}

\subsubsection{TikTok (7/15/2020-3/20/2021)}
\label{tiktok}
TikTok's users report ethical concerns most frequently out of all the apps we analyzed, with 95.9\% (755/787) of reviews disclosing an ethical concern. The most frequent ethical concern is inappropriate content, with a 23.3\% share of users' ethical concerns, with many users reporting violence, including police brutality, filmed suicides, and animal cruelty, as well as pornography. Discrimination is the second most reported ethical concern, at 20.7\% (176/755), 47.7\% (84/176) of these regard some form of racial discrimination. Within 37 of these reports of racism, users also report some other types of discrimination, especially homophobia, transphobia, sexism, anti-Semitism, pedophilia, xenophobia, ableism, and fatphobia. The third highest reported ethical concern is cyberbullying at 17.2\% (130/755) of ethical concerns. These reports of bullying on the app often revolve around inadequacies in the reporting system. These shortcomings include reports of bullying going unaddressed and the 140-character limit for reporting posts being too short.

\subsubsection{Uber (12/7/2013-3/22/2021)}
Uber has the second highest frequency of ethical concerns in their user reviews at 83.21\% (332/399). Safety is their most reported ethical concern at 32.5\% (108/332), with a variety of unsafe situations described, including sexual assault (24/108), dangerous driving (17/108), dangerous route (12/108), or crime and violence (11/108). Next, for the second most occurring ethical concern, accountability, at 20.8\% (69/332), users detail an inadequate reporting system for complaints (17/69), an inability to access their accounts (16/69), and finally getting ignored by customer service (10/69). Lastly, Uber users report scams at a rate of 19.0\% (63/332), with the majority of complaints revolving around Uber misrepresenting pricing, fees, or promotions to manipulate more money out of users. 

\subsubsection{Facebook (7/24/2020-3/22/2021)}
\label{sec:fb}
Overall, Facebook had the third highest rate of ethical concerns, with 72.6\% (246/339) of reviews reporting an ethical concern. Of these ethical concerns, 39.8\% (98/246) relate to censorship, with nearly two-thirds (65/98) referencing political or conservative censorship or narratives surrounding this brand of purported censorship, like biased fact-checking or constitutional rights violations. The rest of the censorship complaints are general or relate to the removal of content in non-US regions'.
Facebook's second most frequently reported ethical concern is identity theft at 14.2\% (35/246). For this concern, users describe Facebook's large volume of publicly available profile data making identity theft easier. Many users are then contacted by fake profiles which seek money or relationships. Lastly,  8.5\% (21/246) of Facebook's ethical concerns relate to privacy, with reports of either corporations or other users using public user data for tracking and spying.
\subsubsection{Instagram (10/20/2020-3/19/2021)}
Instagram's users report ethical concerns at a rate of 47.2\% (111/235). Identity theft is their most frequently reported concern, disclosed at a rate of 39.6\% (44/111). This concern describes catfishing as both identity theft for scamming or harassment and using filters and editing photos to modify one's appearance. Next, users report censorship second most at 9.0\% (10/111), with complaints of right-wing or Hindu content removed. Lastly, users document transparency concerns at a rate of 6.3\% (7/111), with users describing unclear rules regarding bans and deceitfulness concerning data collection. 

\subsubsection{Vinted (7/16/2013-11/15/2020)}
Vinted's users report ethical concerns at a rate of 39.5\% (79/200). Addiction is their most prominent ethical concern, at 51.9\% (41/79); most reviews use the term addiction as a hyperbolic means of describing their enthusiasm for the app. Next, Vinted users report scams at 16.5\% (13/79), with many reviews describing clothes swapping as unequal. The third highest reported ethical concern is accountability; this concern details the flaws in Vinted's customer service system.

\subsubsection{YouTube (1/2/2021-3/22/2021)}
YouTube's users document ethical concerns at a percentage of 27.1\% (87/321) within reviews. Of these, 52.9\% of ethical concerns reported revolve around misinformation; within these reviews users express a desire for the ability to report misinformation and block misinforming users. Inappropriate content is the next highest concern at 44.8\% (39/87); users report a need for age restrictions for ads, comments and thumbnails. Finally, YouTube's users document harmful advertising as their third highest ethical concern at 39.1\% (34/87), with reviews detailing unreportable ads, which contain sexual, drug-related, and gambling content as well as scams. 

 \subsubsection{LinkedIn (8/2/2011-2/20/2021)}
For LinkedIn, users report ethical concerns in app reviews at a rate of 26.7\% (56/210). Their most frequently occurring ethical concern is privacy at 32.1\% (18/56), with reviews mentioning profile data being too readily available to the public and account hacks. Users also report LinkedIn accessing contacts, GPS, and calendar data without permission. Next, users describe discrimination faced on the app at a rate of 23.2\% (13/56); however, users describe ``racism" as targeting both black and white users. Finally,  users detail concerns regarding LinkedIn's transparency at a rate of 17.9\% (10/56). These concerns again describe LinkedIn's attempts to access their device's camera, calendar, and contacts without explaining what will be done with this access. 

\subsubsection{Zoom (4/22/2019-3/4/2021)}
Zoom's users document ethical concerns within their user reviews at a rate of 25.7\% (44/210). Their most commonly occurring ethical concern is privacy, encapsulating a 38.9\% share of reported ethical concerns, with most reviews referring to the ease in which Zoom links can be hacked. Next, users report accessibility issues, with reviews requesting a rotating screen and captions for aid with disabilities. Finally, the third most frequent concern is inappropriate content, with reports of easy imitation, inappropriate virtual backgrounds, and no capability for reporting inappropriate users. 

\subsubsection{Alexa (11/19/2015-3/5/2021)}
Overall, Alexa reports a frequency 20\% (40/200) of ethical concerns among user reviews. Privacy is its main ethical concern, at 62.5\% (25/40). This privacy concern reflects user worries about both the physical device and the app itself. Regarding the app, users are concerned with being forced to install it and asked for permissions for seemingly irrelevant data or capabilities. Similarly, device users report being listened to in physical conversations. Alexa's next-top ethical concern is sustainability at 10\% (4/40); within this concern, users report abnormally high battery drain from the app. The top third concern is addiction with 7.5\% (3/40) of ethical concerns, these concerns detail an overreliance on Alexa as a companion and organization assistant.  

\subsubsection{Google Home (12/3/2013-2/12/2021)}
Google Home's users reported ethical concerns at a rate of 10.5\%(21/200). Privacy is the most frequently reported ethical concern, at 66.6\% (14/21). These privacy concerns revolve around forcing access to their device's GPS, camera, or microphone. Google Home's second most frequent ethical concern is accessibility, at 14.3\% (3/21) of ethical concerns, which all describe a need for captioning. Finally, Google Home's third most reported ethical concern is accountability at 9.5\% (2/21). Reviews detail leaving complaints that receive no response from customer service and being unable to find options for customer service.

\subsection{Ethical Concerns Over Time (RQ1)}
\begin{figure}[h!]
    \centering
    \caption{Top 5 Ethical Concerns' Weekly Frequency from 1/1/21-3/21/21}
    \centerline{\includegraphics[width=.85\textwidth]{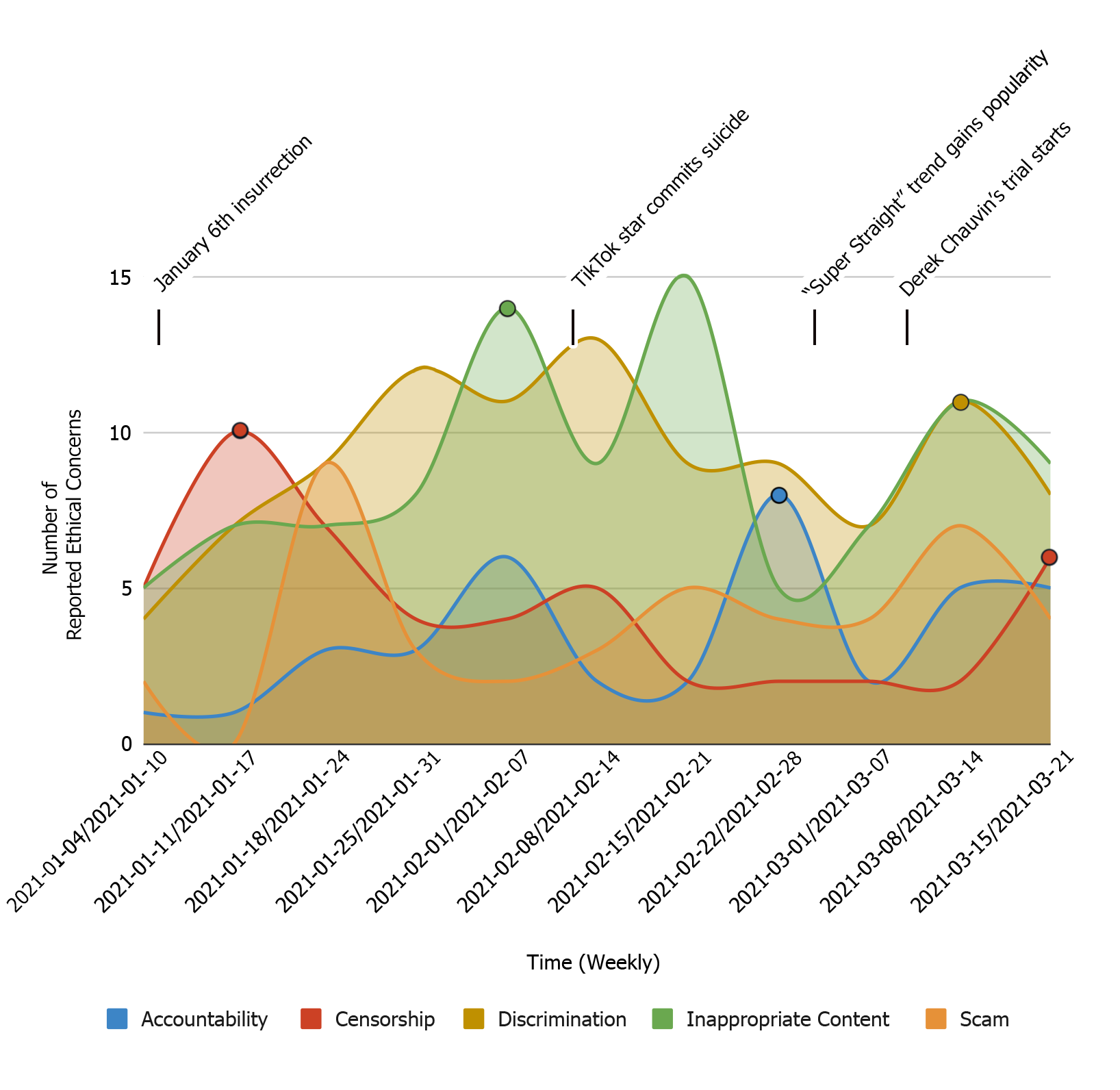}}
    \label{fig:ethictime}
\end{figure}

The ethical concern frequency per year is normalized by the total amount of app reviews collected per year. Within this section, we examine all data peaks for the top 5 ethical concerns that appear most frequently, as seen in Figure~\ref{fig:ethictime}. We report those which show some relevant trend. We analyze weekly frequencies from the start of 2021, as the latest start date of the analyzed applications is January 2, 2021 (see Table~\ref{table:time}).  

\subsubsection{Accountability}
\label{sec:peak1}
\textbf{2/22-2/28.}
In general, posts detail reporting bad content and not seeing change. More specifically, nearly 40\% (3/8) user app reviews are about vulgar content on TikTok. This peak occurs two weeks after prominent TikTok star commited suicide. One of the posts mentions the suicide. The TikTok star gained prominence after a video of her arguing with her drug addict mom went viral~\citep{tiktokstar}.

\subsubsection{Censorship}
\label{sec:peak2}
\textbf{ 1/11-1/17.}
These posts all regard right-wing censorship on YouTube (6), TikTok (2), and Facebook (2). After the January 6th insurrection at the US Capitol, all of these platforms increased regulation of content relating to Trump. YouTube barred Trump from uploading new videos or livestreams and disabled comments on his videos, removed some of his videos~\citep{Pitofsky_2023}. TikTok’s ``integrity and authenticity” proactive removal rate increased by nearly ten percent from July-December 2020 to Jan-March 2021~\citep{TikTok_2022}. This ``integrity and authenticity” proactive removal rate may be connected to TikTok removing misinformation relating to the 2020 US Presidential election and Trump's loss, which played a role in engendering the insurrection. Facebook banned Trump’s Facebook for at least two weeks. Facebook also reported that they actively removed content relating to praise or encouragement of storming the Capitol or reports of related illegal or violent gatherings in the US~\citep{Rosen_Bickert_2021}.

\noindent \textbf{3/15-3/21.}
The majority of posts (4/6) relate to left ideologies being promoted or conservative content being removed or demoted. The trial of Derek Chauvin started one week before this peak. Derek Chauvin was tried for the murder of George Floyd, whose death catalyzed momentum for the Black Lives Matter protests of 2020. Reportedly, social media platforms like Facebook, YouTube and Twitter all `braced' for the verdict of the trial by increasing their efforts of impeding the spread of misinformation and harmful content~\citep{Pitofsky_2023}.

\subsubsection{Discrimination}
\textbf{3/8-3/14.}
This discrimination peak's posts start to mention a new trend: `superstraight'. This trend started in late February and began to gain popularity over the next few weeks, spreading to other platforms at the beginning of March~\citep{Asarch_2021}. `Superstraight' is defined as a sexuality where one is romantically or sexually attracted to only cisgender people of the opposite sex. Nearly 40\% (4/11) of posts in this peak reference being discriminated against for being superstraight.

\subsubsection{Inappropriate Content}
\textbf{2/1-2/7.}
Around 43\% (6/14) of posts in this peak regard nudity and sexual content. Two of the posts seem to reference the same event, where a man touching his genitals online was left on TikTok for 24 hours, despite many reports of the offending post. 

\subsection{Characteristics of Ethical Concerns (RQ2)}
\begin{figure}[h!]
    \centering
    \caption{Average User Rating per Ethical Concern}

    \centerline{\includegraphics[width=.75\textwidth]{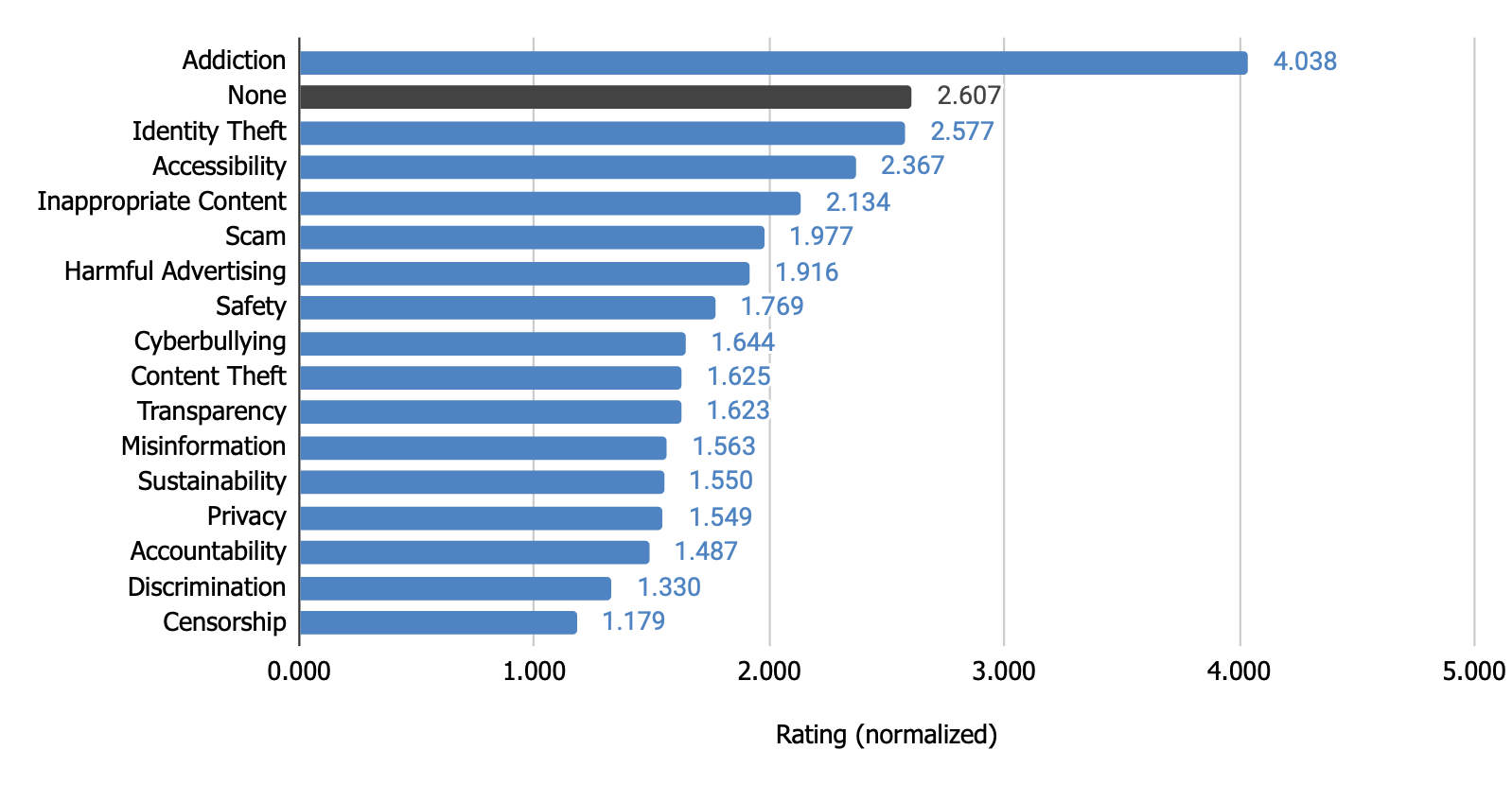}}
    \label{fig:rating_per_eth}
\end{figure}

For all reported characteristics, each ethical concern's average was calculated per application then divided by the total amount of reviews for that application. These values were then summed across all applications. This procedure scales the characteristics so each application is balanced equally. 
\label{sec:rating}
 User reviews mentioning ethical concerns had an average normalized rating of 1.896, with a median of 1.635 and a standard deviation of .679. In comparison, users' reviews without ethical concerns had an average rating of 2.607. Of the ethical concerns, addiction had the highest rating, 4.038, while censorship had the lowest rating at 1.179. Figure~\ref{fig:rating_per_eth} shows the ratings for each ethical concern.

\begin{figure}
    \centering
    \caption{Average Word Count per Ethical Concern}
    \centerline{\includegraphics[width=.75\textwidth]{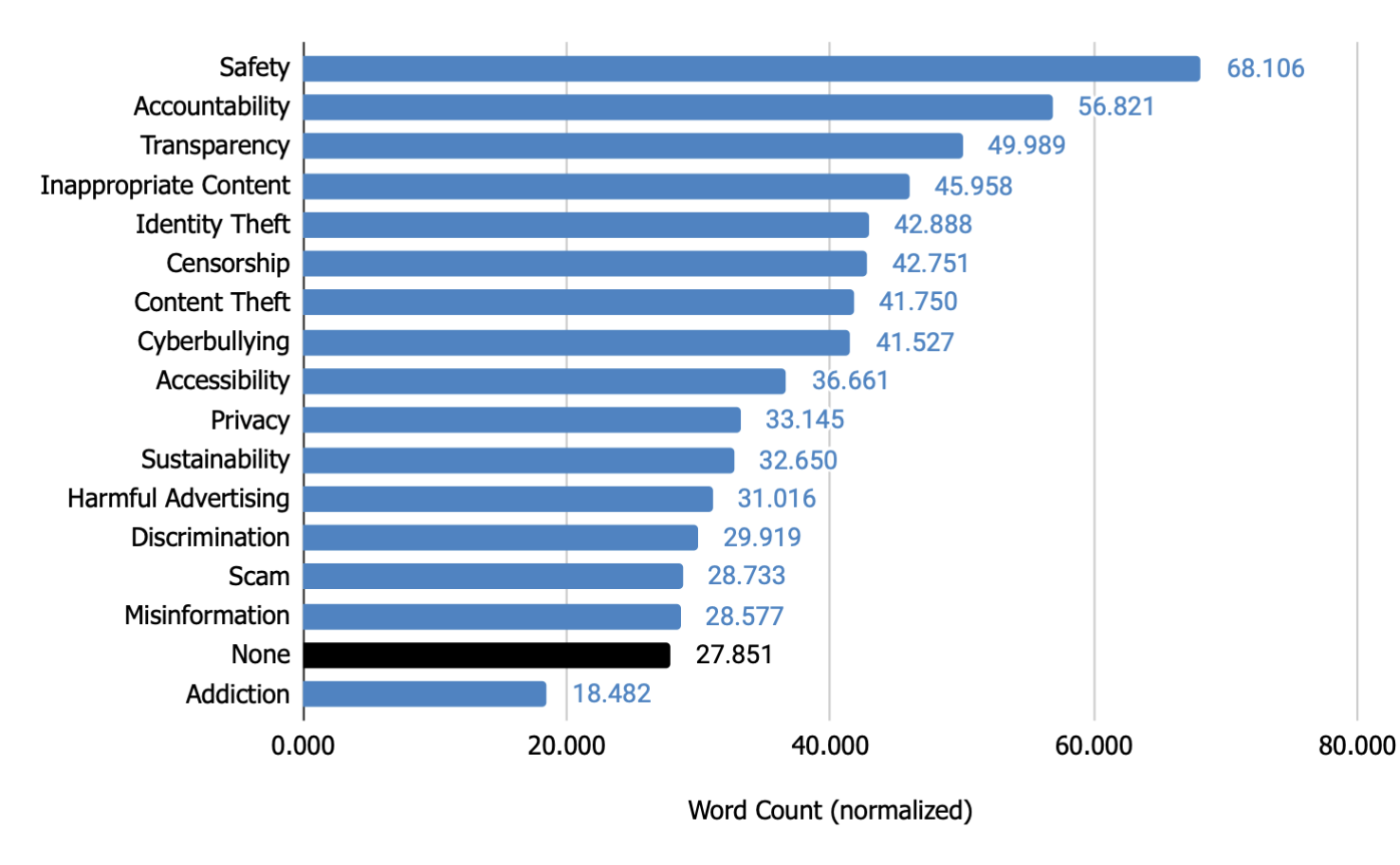}}
    \label{fig:wordcount}
\end{figure}
Overall, the \textbf{normalized word count} of reviews mentioning an ethical concern had an average of $39.311$ (median=$39.094$ and standard deviation=$12.212$), while those without had an average word count of $27.851$. The ethical concern with the highest word count was safety with an average of $68.106$, while addiction was the ethical concern with the lowest word count, with an average of $18.482$. 

\begin{figure}
    \centering
        \caption{Average Up-vote Count per Ethical Concern }
    \centerline{\includegraphics[width=.75\textwidth]{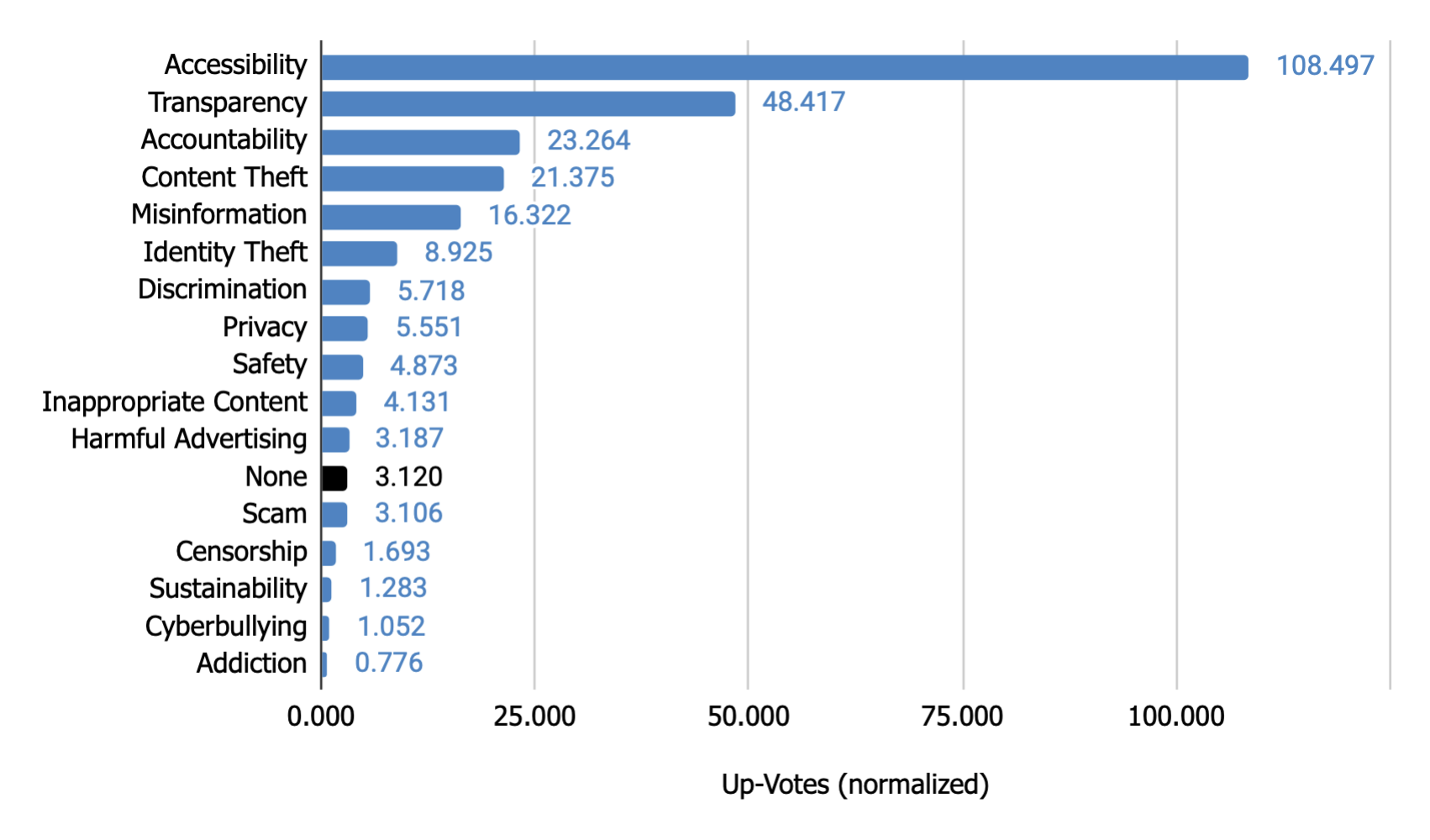}}
    \label{fig:thumbsupcount}
\end{figure}
 The reviews labeled with an ethical concern had an average normalized up-vote count of $16.136$ (median=$5.212$, standard deviation=$27.598$) while reviews without ethical concerns had an average up-vote count of $3.120$. Accessibility is the ethical concern with the highest average up-vote count, $108.497$, while addiction is the ethical concern with the least average up-vote count, $0.776$.

 

\subsection{Automated Analysis of Ethical Concerns (RQ3)}
\textbf{Binary Classification of Ethical Concerns}
These classifiers categorize each review as mentioning an ethical concern or not. In total, 62\% of the filtered reviews mentioned an ethical concern, while 38\% did not.  
The performance scores of the five evaluated classifiers with and without SMOTE (see Section \ref{sec:smote}) are similar, with all models having an F1-score of at least 0.87. Table~\ref{table:results_bin} shows the main results for each classifier. 
\begin{table}[h]
\caption{Binary Classification Results for Ethical Concern Presence}
\centering
\begin{tabular}{p{0.2\textwidth} c c c}
\hline
\textbf{Model}& \textbf{Precision}& \textbf{Recall}  & \textbf{F1-score}  \\ \hline
\rowcolor{Gray}\textit{RF} &  0.883 & 0.863 & 0.872 \\ 
\textit{RF with SMOTE} & 0.874 & 0.878 & 0.875   \\ 
\rowcolor{Gray}\textit{SVM} & 0.897 & 0.879 & 0.887   \\ 
\textit{SVM with SMOTE} & 0.867 & 0.884 & 0.873   \\ 
\rowcolor{Gray}\textit{MNB} & 0.889 & \textbf{0.898} & \textbf{0.893 }  \\ 
\textit{MNB with SMOTE} & 0.889 & 0.891 & 0.887   \\ 
\rowcolor{Gray}\textit{LR} & \textbf{0.898} & 0.878 & 0.886   \\ 
\textit{LR with SMOTE} & 0.879 & 0.895 & 0.885   \\ 
\rowcolor{Gray}\textit{MLP} & 0.892 & 0.877 & 0.883  \\ 
\textit{MLP with SMOTE} & 0.872 & 0.879  & 0.874  \\ \hline
\rowcolor{Gray}\textit{BERT} & 0.94 & 0.94  & 0.94  \\ \hline

\end{tabular}
\label{table:results_bin}
\end{table}

\textbf{Binary Classification of Internal and External Ethical Concerns}
For some reviews mentioning ethical concerns (3/583), it was difficult to assign an \textit{internal/external} label because both labels applied or annotators were unsure about their labeling. Therefore, we left these reviews out of the data we used to train and test \textit{internal/external} models. With the inclusion of only our potential \textit{external} categories, detailed in \ref{sec:inex}, we had a dataset that contained $583$ reviews labeled as \textit{internal} and 143 reviews labeled as \textit{external}. We used this data for the training and testing of the models. 

Table~\ref{table:results_bin_inex} shows the performance scores for the classifiers that assigned \textit{internal} and \textit{external} labels to reviews. All the models used for this classification performed almost equally well. The highest F1-score was achieved by the Logistic Regression, $0.957$, whereas the Random Forest with SMOTE achieved the lowest F1-score, $0.921$.

\begin{table}[h]
\caption{Binary Classification Results for Internal and External Labels}
\centering
\begin{tabular}{p{0.15\textwidth}  c c c}
\hline
 \textbf{Model} & \textbf{Precision} & \textbf{Recall} &\textbf{F1-score} \\ \hline
\rowcolor{Gray}\textit{RF} & 0.944 & 0.922 & 0.929  \\ 
\textit{RF with SMOTE} & 0.942 & 0.913 & 0.921   \\ 
\rowcolor{Gray}\textit{SVM} & 0.959 & 0.924 & 0.935  \\ 
\textit{SVM with SMOTE} & 0.955 & 0.938 & 0.944  \\ 
\rowcolor{Gray}\textit{MNB}& 0.920 & 0.937 & 0.924  \\ 
\textit{MNB with SMOTE}& 0.918 & 0.940 & 0.923  \\ 
\rowcolor{Gray}\textit{LR}& \textbf{0.973} & 0.949 & \textbf{0.957}  \\ 
\textit{LR with SMOTE}& 0.962 & 0.955 & 0.956  \\ 
\rowcolor{Gray}\textit{MLP}& 0.964 & 0.943 & 0.951  \\ 
\textit{MLP with SMOTE}& 0.959 & \textbf{0.958} & \textbf{0.957}  \\ \hline
\rowcolor{Gray}\textit{BERT}& 0.97 & 0.97 & 0.97  \\ 

\end{tabular}
\label{table:results_bin_inex}

\end{table}

\textbf{Multi-Class Classification of Ethical Concerns} The models trained in this step classified the reviews mentioning an ethical concern into the categories shown in Table~\ref{table:taxonomy}, excluding harmful advertising, accessibility, sustainability, and content theft, due to lack of samples, as shown in Figure~\ref{fig:distr}.
Table~\ref{table:results_eth} shows the results of the multi-class predictive models with macro-averaging. From these models, the Random Forest with SMOTE  and Logistic Regression with SMOTE both achieved the highest  F1-score, $0.83$. Within the Logistic Regression with SMOTE model, the categories with the highest F1-scores are addiction (0.95), identity theft (0.92), and privacy (0.88) while the lowest are accountability (0.79), inappropriate content (0.77), and discrimination (0.77).

\begin{table}[hbt!]
\caption{Multi-Class Classification Results}
\centering
\begin{tabular}{p{0.15\textwidth} c c c }

\hline
 \textbf{Model} & \textbf{Precision} & \textbf{Recall} & \textbf{F1-score} \\
 & \textbf{macro} & \textbf{macro} & \textbf{macro} \\ \hline
\rowcolor{Gray}\textit{RF}& 0.844 & 0.812 & 0.817  \\ 
\textit{RF with SMOTE}& 0.839 & 0.812  & 0.816  \\ 
\rowcolor{Gray}\textit{SVM}& 0.825 & 0.796 & 0.803  \\ 
\textit{SVM with SMOTE}& 0.818 & 0.800 & 0.801  \\ 
\rowcolor{Gray}\textit{MNB}& 0.837 & 0.804 & 0.811   \\ 
\textit{MNB with SMOTE} & 0.821 & 0.808 &  0.808  \\ 
\rowcolor{Gray}\textit{LR}  & \textbf{0.850} & \textbf{0.821} & \textbf{0.830 } \\ 
\textit{LR with SMOTE}& 0.835& \textbf{0.821} & 0.822   \\ 
\rowcolor{Gray}\textit{MLP}& 0.834 & 0.805 & 0.812  \\ 
\textit{MLP with SMOTE}& 0.817 & 0.800 & 0.802  \\ \hline
\rowcolor{Gray}\textit{BERT}& 0.82 & 0.81 & 0.82  \\ 

\end{tabular}
\label{table:results_eth}
\end{table}

\label{sec:res_automatic}
We explored the use of traditional machine learning methods for the automatic extraction and classification of ethical concerns. We developed three types of predictive models: (1) binary classifiers for detecting the mention of ethical concerns, (2) binary classifiers for detecting whether an ethical concern is \textit{internal} or \textit{external}, (3) multi-class classifier categorizing the ethical concerns shown in Table~\ref{table:taxonomy}.

\section{Discussion} 
\label{sec:discussion}
Our results demonstrate that \textit{(RQ1)} on the Google Play Store users' ethical concerns center around inappropriate content, accountability and discrimination, \textit{(RQ2)} user reviews with ethical concerns are longer, more popular and have lower ratings, and \textit{(RQ3)} there is high automation potential for finding ethical concerns within filtered data.

\subsection{Types of ethical concerns present in user reviews \textbf{(RQ1)}} Censorship, identity theft, and safety were the most recurring ethical concerns in our sample, while accessibility and sustainability appeared the least. In addition, we found that ethical concerns occur in 61.9\% (1919/3101) of our sample; this compares to previous research which finds that 26.5\% of reviews contain expressions of human value violations~\citep{obie2021first}; we find it likely that our keyword filtering technique inflated our sample's frequency by over 30 percent. We also found that multiple ethical concerns occur in only 11\% of our sample; this result contrasts to the work of Shams et al.~\citep{shams2020society} which found that multiple human value violations are more predominant than single ones. This work used the Schwartz value framework. This framework is not directly related to software applications and thus we posit the ambiguity of their categories leads to less clear delineation between categories (see Section~\ref{sec:related_work}). The applications considered in this study differed in the degree to which ethical concerns were mentioned the most; however, we could find some similarities between application domains. We found that the ethical concerns discrimination, inappropriate content, censorship, and privacy were mentioned often for social media applications. AI assistants and Zoom had privacy issues as their most reported ethical concern, and sharing economy applications had accountability and scam as one of their most mentioned ethical concerns. 

Furthermore, we discovered some general ethical patterns by analyzing ethical concerns per application (see Section \ref{sec:qa}). For TikTok, Uber, Vinted, YouTube, Zoom, and Google Home, thus multiple domains, users describe systems built to dismiss users' concerns, clearly signaling an ethical vulnerability. These complaints detail reporting systems that are inadequately designed to handle misinformation, cyberbullying, safety, discrimination, scams, and other ethical concerns. Second, users disclose obfuscation around using their data and devices within applications across domains, including LinkedIn, Google Home, Alexa, and Facebook. These general trends signal that software users not only highly value clarity surrounding how applications treat them and their data, but are astutely aware of implicit attempts to disregard their concerns. Therefore, when building software systems, all practitioners should consider how they can build adequate reporting systems into myriad design features, to allow users to give more fine-grained and targeted feedback on content and experiences. 

Our analysis of ethical concerns over time provides further evidence that the content and frequency of users' app store reviews can be tied to real world events, like the US presidential election. This finding clarifies a critical connection between politics and users' ethical concerns, signalling a need for development teams to be aware of how myriad world events may affect users' requirements. Additionally, by considering temporal shifts in user ethical feedback, practitioners can more easily evaluate the effectiveness of ethical software interventions. 

Most of the ethical concerns reported in user reviews occurred while on the application (92\%, \textit{internal}), while only 8\% happened physically \textit{(external)}. The three categories that were most often associated with the \textit{external} label were safety, sustainability, and scam. These three ethical categories strongly relate to the service the application provides where the service either caused a dangerous situation, environmental issues, or financial losses for end-users. The critical nature of these claims exemplifies the importance of the internal/external classification for ethical concerns. By tagging concerns as external, practitioners are given are given extra information regarding the extent and impact of the ethical concern. This tag therefore can be used by practitioners as a metric within prioritization systems.

\subsection{Characteristics of user reviews that mention ethical concerns \textbf{(RQ2)}} Reviews mentioning ethical concerns had a lower average rating (1.90) than those that do not mention ethical concerns (2.61). The average rating among reviews mentioning ethical concerns is also considerably lower than the average rating found in previous work (4.13, 3.25, 3.51) (\cite{Dennis2013, Guzman2018,fischer2021does}). Similar to our results, Pagano and Maalej found that privacy and ethical issues had a negative effect on ratings~\citep{Dennis2013}. The aforementioned research also found that negative review text tends to lead to lower ratings~\citep{Dennis2013}. We also found that most reviews that mention ethical concerns were longer than those that do not (39.31 vs. $27.85$ words); and that it is also higher than those reported in previous work analyzing general user reviews about software (18) (\cite{Guzman2018}). Similarly, reviews mentioning ethical concerns were, on average, more upvoted than those not mentioning ethical concerns. These results might not be surprising as reviews reporting issues could tend to have lower ratings, need more words for their explanation, and create more polarization, expressed in the forms of up-votes. While we compared amongst reviews that did not contain ethical concerns, it would be interesting to compare exclusively against reviews reporting technical malfunctioning, i.e., bug reports, as they are another type of review describing issues. 

\subsection{Automation potential \textbf{(RQ3)}}\label{sec:discussionAP} The large number of reviews and the relative sparsity of those mentioning ethical concerns call for the use of automatic analysis techniques. In this regard, our experiments using traditional machine learning classifiers are promising. The classifiers detected ethical concerns with an F1 score of 0.89 and classified ethical concerns as internal or external with an F1 score of 0.96. Furthermore, they could categorize them into our set of ethical concerns with an F1 score of 0.83. However, these results must be treated with caution because we collected all data using keyword filtering (see Section~\ref{sec:limitations}). 

As machine learning tasks with small datasets and subjective criteria infrequently achieve results as high as ours, we will attempt to explain our models' success. We find that this success is likely to be attributed to our keyword filtering process. By filtering the data set by keywords, we reduce the vocabulary size from 835,551 to 14,755 words. We also increased the average review's length from 5.54 to 34.30 words, giving greater context to each review. Finally, we posit that this restricted the overall conversational space; by limiting the discussed topics, we introduce our model to less noise, making it more likely to revisit similar material. To investigate this more deeply, we examine the features of the Logistic Regression binary classification model (with SMOTE), as it is a high-performing model with accessible features and feature importances. Overall, filtering keywords occurred in 26.86\% of the feature importances used by the model. Of these keyword-features, 65\%(100/156) indicated the review being an ethical concern. Therefore, we assume that our model would likely not perform as well our full Google Play Store dataset, but would perform highly on the around 330,000 reviews in our filtered dataset. This high performance would allow an accurate classification of these samples and the potential for a further detailed analysis of trends within these ethical concerns. While the nature of our classification task is different to previous work---classifying for example feature request or bug reports in app store reviews or social media~\citep{maalej2015r,guzman2015r, villarroel2016release,guzman2017little}---the results are comparable.

\subsection{Implications} This work is a step towards systematizing the consideration of ethical concerns during software development and evolution. The set of\textit{ user-informed ethical concerns} found in our study can serve as a \textit{checklist} for practitioners of concerns they should take into account when designing software and when 'testing' their absence before software release and during evolution. We stress that this set of user-informed ethical concerns did not exist prior to our study. Similarly, researchers could use the proposed set to analyze the existing ethical concerns of mobile software applications. 

Since we found that end-users report ethical concerns in user reviews, practitioners and researchers could use this specific user feedback channel and automated tools to analyze which ethical concerns remain unfulfilled in their software and are most pressing for their users.

The internal and external classification provides the developer with more information regarding the extent of the issue. Once an ethical concern becomes external, it may become a higher priority, as it may present a physical threat to the user(s). In these situations, there is potential risk for serious harm for the user, litigation, and damage to the platform’s reputation. As such, it may be necessary to involve other departments (legal, public relations) while crafting a solution.  It also may be critical for software professionals to contact proper authorities or otherwise quickly intervene to ensure the safety of their users (consider situations regarding sexual assault, self-harm, or other acts of physical violence). It could also therefore provide practitioners with information regarding the scheduling of various tasks.

The quantitative results regarding rating, word count, and popularity measures can be used as features to machine learning classifiers to help practitioners more easily automatically identify and categorize user reviews with ethical concerns. These quantitative characteristics also reveal the nature of ethical concern reviews, polarizing, lengthy, and negative. 

Our qualitative findings provide specific, actionable feedback for researchers and developers to enhance their ethical practices and for companies to gain a competitive edge in the market. For example, through our analysis, we have identified key areas of focus across domains, such as the urgent need for improved reporting systems and greater transparency around permission requests. The latter result matches previous work, which finds that app reviews containing privacy issues have permission requests as a dominant theme~\citep{nema2022analyzing}. Our analysis on specific applications also gives the concerned companies and their competitors insight into which ethical concerns are present among their user base. In addition, practitioners and researchers can use our dataset, models, and set of ethical concerns to investigate different types of ethical concerns, create time series analyses, and investigate the concerns of different populations and software application domains.

\textbf{Suggestions for Practitioners. }
Based on our results, we further detail our two suggestions for practitioners: (1) improved reporting systems and (2) greater transparency around permission requests. In our work, users report insufficient character limits for reporting content, opacity during complaint processes, unclear rules regarding bans, and unreportable interface elements, including ads, users, and comments. As a solution, practitioners should increase character lengths for reports to ensure users can report all user forms of user-generated content. Indeed, previous research regarding user preferences on content moderation finds similar results: users want increased clarity of interface elements' definitions, incorporated context, and appropriate levels of granularity for report categories~\citep{jhaver2023personalizing}. These improvements should include providing examples of content that fit into each category. While our users did not identify report categories as an issue, the chosen categories of media platforms define vocabularies by which users can report content; practitioners choose the types and granularities of categories of inappropriate content. This ability \textit{determines} users' reporting capacity~\citep{crawford2016flag}. As such, categories and their granularities must also be reviewed - doing so may reduce the need for increased character limits in reviews, streamlining the process. Some platforms, like Facebook, have created a user-facing dashboard that shows the status of user reports; we suggest this type of feature to increase transparency surrounding user bans and reports.

Second, users desire increased transparency surrounding permission requests; often, they do not understand why the application is requesting permissions, or it may seem like the requested permission is unnecessary to the app's function. Previous research suggests that permission issues are common on Android apps, especially \textit{over-permissions}, where too many or superfluous permissions are requested~\citep{mujahid2018studying,scoccia2019permission}. As such, applications should review their code relating to permissions to ensure minimal permissions are requested and, when they are requested, the purpose of this additional data acquisition is made explicitly clear to the user.

\section{Threats to Validity}\label{sec:limitations}

\subsection{Construct Validity}
We treated the problem of classifying ethical concerns into the categories of our set of ethical concerns (see Table~\ref{table:taxonomy}) as a multiclass problem, although there were cases where more than one ethical concern was present in the review. However, multiple labels were only present in 11\% of the reviews. Additionally, the annotators assigned the most predominant ethical concern in the review as the first label.

 A flaw of our data collection arises from differences in application popularity; due to the varied volumes of reviews, the time period of user reviews for each application is different. For example, the LinkedIn user review history in our work covers nine years, while we only needed to gather three months of Instagram user reviews to get a million reviews. This lack of temporal uniformity may have influenced the ethical concerns mentioned in the application, as there is an interplay between ethical concerns in the physical and online world. 

\subsection{Internal Validity}
Although we took considerable care in the creation of our user-informed set of ethical concerns, this set is non-exhaustive and could lack ethical concerns unrelated to those we used from Wright~(\cite{wright2011framework}) or found in user reviews.  This threat was the motivation behind performing nine trial rounds of annotation, even though annotators reached a substantial agreement (0.63) in the fifth round. Indeed, after four annotation trials, the annotators found no new concerns (all nine new ethical concerns were found in the initial trials). 

Although the annotators carefully annotated the reviews according to the guideline, annotating ethical concerns can be a subjective matter. It was, in some cases, challenging to see whether the application caused an ethical concern or whether the user inflicted the problem. For example, when deleting their data accidentally and blaming the application for it. To simplify this, the annotators assumed the user's truth, which the results reflect. 

When labelling for addiction, we labelled reviews mentioning overuse of the app, even if no negative consequences were detailed in the review. We made this decision due to the complex nature of the ethical concern. In order for some phenomenon to be addictive, it must have a highly positive aspect to draw the user. Due to the critical nature of this ethical concern, and clear evidence of the negative health effects on users~\citep{hou2019social, huang2022meta, paakkari2021problematic}, we consider it crucial to ensure these reviews are considered in our analysis. Nevertheless, there is a potential overlabelling of addiction as a concern due to a lack of provided information from users in app review due to its vernacular use as hyperbole (‘I’m addicted’ sometimes just means ‘I love it’). This may not just inflate frequency but the user review score.

\subsection{External Validity}

Although keyword filtering proved an effective technique for finding relevant reviews on ethical concerns, our dataset could lack reviews using vocabulary outside of our keyword list. We addressed this potential limitation by (1) expanding the initial list with other relevant words found in the nine annotation trials (new keywords were not found after the fourth round) and (2) including synonyms of the keywords from a well-known thesaurus. However, this limitation could have an effect on the diversity of the collected dataset, the reported frequencies of ethical concerns, and on the performance of the classifiers (see Section~\ref{sec:discussionAP}). Because of this fact, we cannot claim that the most predominant ethical concerns found in our study are, in fact, the ones that are most mentioned in the whole Google Play Store ecosystem or ensure that our classifiers will work equally well in unfiltered datasets. Nevertheless, snowballing for keywords has been used by previous research when analyzing (single) ethical concerns in user feedback to overcome sparse data e.g.,~(\cite{tushev2020digital, obie2021first}).

Tizard et al.,~(\cite{tizard_rietz_blincoe_2020}) found that, across all channels, men give more feedback than women, and the most common age group that gave feedback was between 35 and 45. Therefore, it is critical to note that our results are likely to represent the perspective of this demographic group. For example, within the censorship reports on Facebook discussed in Section~\ref{sec:fb}, most reports relate to conservative political affiliations, which older men tend to have (\cite{saad_2022}). Further research should target the opinions of other demographic groups to better represent all users' interests, rather than those whose opinions are repeatedly centered (\cite{costanza-chock_2020}).

We collected data from nine different applications from four different domains. All applications are widespread, reach different target audiences, and have different end goals. Nevertheless, further studies analyzing additional applications from missing domains and with less popularity need to be conducted to conclude if our results generalize. Furthermore, additional studies should be performed to analyze if the results hold on user feedback written in other app stores and user feedback channels. 


\section{Related Work}
\label{sec:related_work}
Research on user feedback for software engineering has grown considerably in recent years. 
This section discusses related work in using user feedback to develop software applications, approaches to automatically mine user feedback, as well as previous work examining ethical concerns expressed in user feedback.

\textit{User Feedback and Software Evolution.}
Previous work found that user feedback is essential for software quality and identifying areas of improvement (\cite{Pagano2013}). With the rise of mobile applications and social media, research proposed to elicit feedback from crowds of geographically distributed users (\cite{groen2017crowd}) and called for the mass participation of software users during different stages of software development (\cite{Johann2015}).
Pagano and Maalej~\citep{Dennis2013}, and Hoon (\cite{hoon2013analysis}) were among the first to study user feedback in app stores.
They performed exploratory studies and found that this platform contains valuable information for software evolution, such as bug reports or feature requests. On a similar line, more recent research has found that software practitioners often use app store reviews for eliciting requirements and software evolution~\citep{van2021role,li2023unveiling}. The results of this work motivated the use of app reviews in our study.  

\textit{Mining User Feedback.} Among the most studied platforms for automatically processing user feedback are app stores. Martin (\cite{martin2016survey}) described a survey of the most relevant work in the area, we refer to them for a more comprehensive analysis of the area. 
Previous work has proposed approaches for classifying, e.g.,(\cite{guzman2015r,maalej2015r,panichella2015r}), grouping, e.g.,(\cite{Chen2014,villarroel2016release,disorbo2016,GalvisCarreno2013,Iacob2013}) and prioritizing, e.g., (\cite{Chen2014,villarroel2016release,disorbo2016, kifetew2021automating}) user feedback, as well as for extracting software features mentioned in the feedback (\cite{Guzman2014,gu2015parts}) and linking it to other artifacts (\cite{palomba2017recommending}). In our work, we use predictive models for detecting and classifying ethical concerns, taking similar preprocessing steps as those described in the aforementioned previous work~\citep{guzman2015r,guzman2017little,williams2017mining,maalej2015r,panichella2015r}. 

\textit{Ethical Concerns in User Feedback.}
Studies on ethical concerns mentioned in user feedback are still sparse and have mostly focused on single ethical concerns. The most studied ethical concern is privacy (\cite{besmer2020investigating,khalid2014mobile,li2022narratives,nema2022analyzing}), but other concerns, for example discrimination~\cite{tushev2020digital} have also been studied. While there are studies that have analyzed specific ethical concerns from an end-user perspective, we are the first to look beyond singular ethical concerns, with the goal of contributing to the systematic consideration of ethical concerns during software development and evolution. Conversely, both Shams et al., (\cite{shams2020society}) and Obie et al., (\cite{obie2021first}) studied \textit{human values violations} within app store reviews via the Schwartz theory of human values (\cite{schwartz2012overview}). However, this theory was created for the psychology field, so its values do not always align closely with software (e.g., ``humble" and ``world of beauty").  This work is the first to analyze different types of ethical concerns about software reported by end-users. By contrast, our ethical concerns derive from software-related literature~\cite{wright2011framework} and app reviews that directly relate to software applications. 
\section{Conclusion}
In this work, we analyzed ethical concerns in user reviews from popular software applications.  We found that users report ethical concerns, albeit sparsely. The most common ethical concerns reported by users include censorship, identity theft, and safety. Our analysis of ethical concerns over time provides further evidence that their mention and frequency can be tied to real world events.  Additionally, we found that reviews mentioning ethical concerns are lengthier, more negative in rating, and more up-voted than reviews not mentioning ethical concerns.  The large number of reviews and relative sparsity of reviews mentioning ethical concerns calls for the use of automatic processing techniques for its filtering and classification. In this respect, the experiments conducted in this study, using traditional machine learning and deep learning, are promising. Our results contribute to a more systemic consideration of ethical concerns during software evolution. 

\section{Data Availability Statement}
To encourage replication and further research, we make our data and models publicly available\footnote{\label{rep1}https://doi.org/10.5281/zenodo.7755689}.

\noindent The authors declare that they have no conflict of interest.

\bibliography{citations}










\end{document}